\documentclass[12pt]{iopart}

\usepackage{graphicx}
\usepackage{color}
\usepackage[english]{babel}
\usepackage[utf8]{inputenc}
\begin{document}

\title{Analyzing the dynamics of a \textit{yoyo} using a smartphone gyroscope sensor}

\author{Isabel Salinas$^1$, Martin Monteiro$^2$, Arturo C. Marti$^3$, Juan A. Monsoriu$^1$}

\address{$^1$
Universitat Politecnica de Valencia, Valencia, Spain}

\address{$^2$ Univesidad ORT Uruguay}
\address{$^3$ Facultad de Ciencias, Universidad de la Rep\'{u}blica, Uruguay}

\ead{isalinas@fis.upv.es}
\ead{monteiro@ort.edu.uy}
\ead{marti@fisica.edu.uy}
\ead{jmonsori@fis.upv.es}

\maketitle

In this article, the dynamics of a traditional toy, the yoyo, is investigated theoretically and experimentally using smartphone' sensors. In particular, using the gyroscope the angular velocity is measured. The experimental results are complemented thanks to a digital video analysis. The concordance between theoretical and experimental results is discussed. As the yoyo is a ubiquitous, simple and traditional toy this simple proposal could encourage students to experiment with everyday objects and modern technologies.

The usage of toys to teach physics is an interesting approach to promote engagement and creativity \cite{guemez2009toys}. Traditionally, toys have been widely used in qualitative demonstrations. However,  it is frequently difficult to extract quantitative results in physics experiments involving toys. One possible strategy to address this difficulty is the use of smartphone sensors. 
Among the spectrum of sensors available the gyroscope (also known as angular velocity sensor),
clearly well suited to experiment in situations involving rotational dynamics, has received comparatively less attention  \cite{Monteiro2014angular,Monteiro2014exploring,MONTEIRO2015,monteiro2014rotational,0031-9120-54-1-015003,patrinopoulos2015angular,porn2016interactive}.
Here, we focus on smartphone gyroscope to analyze the dynamics of a yoyo.

\begin{figure}[h]
\begin{center}
\includegraphics[width=0.67\textwidth]{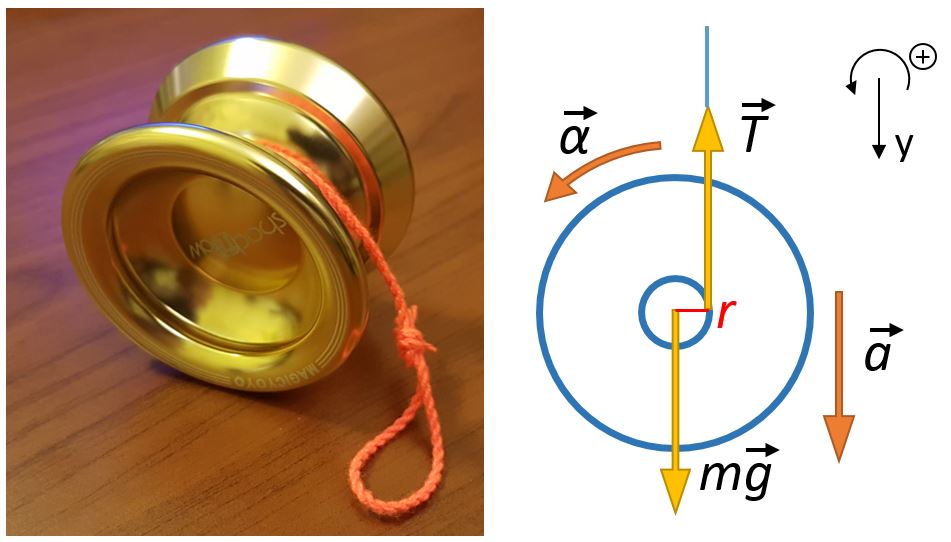}
\caption{A yoyo (left panel) and a free body diagram (right).}
\label{fig-yoyotradicional}
\end{center}
\end{figure}

The yoyo is a traditional toy whose origins can be traced back at least to ancient Greece \cite{yoyohistory} (Fig.~\ref{fig-yoyotradicional}). It is made up of two disks joined by an axle and a string wound around the axle. The players holds one end of the string with their hand by inserting one finger into a small knot and executes several plays. In the simplest version, the player
lets the toy to fall by the effect of the gravity, spinning and unwinding the string.  When the yoyo reaches the lowest position, it bounces back and starts climbing as the string is wound again in the axle. To counteract the friction effects the player applies small pushes, up or down, when the yoyo is climbing or falling respectively.

\textbf{The physics of the yoyo.} We consider a yoyo, with mass $m$, hanging from a string and moving only in the vertical direction. The string is considered as weightless and inextensible. The applied forces are the tension of the string, $\vec{T}$, and the weight, $m\vec{g}$ as indicated in Fig.~\ref{fig-yoyotradicional} (right panel) both acting in the vertical direction. Assuming the $y$-axis oriented downwards,
the second Newton law can be written as
\begin{equation}
  ma_y = mg - T 
  \label{eq:new}
\end{equation}
where $a_y$ is the vertical acceleration of the center of mass.

We assume that the tension is applied upwards at a perpendicular distance which coincides with the radius of the inner cylinder, $r$, as shown in Fig.~\ref{fig-yoyotradicional}. 
Taking counterclockwise rotations with positive sign,
the rotational version of  Newton's second law in the center of mass can be expressed as
\begin{equation}
I \alpha = \pm T r
\end{equation}
where $I$ is the moment of inertia and  $\alpha$ is the angular acceleration,
both magnitudes relatives to an axis through the center of mass and perpendicular to the plane of the
yoyo. The  positive (negative) sign in the previous equation corresponds to the yoyo hanging on the left (right) side of the string. The  non-slipping condition of the yoyo, relative to the string,  relates the vertical  and the angular acceleration
\begin{equation}
a_y =  \alpha  r.
\label{eq:non}
\end{equation}
The angular acceleration can be obtained solving Eqs.~\ref{eq:new}-\ref{eq:non},
\begin{equation}
\alpha = \pm g \frac{r}{r^2+I/m}.
\label{eq:alpha}
\end{equation}
Equation~\ref{eq:alpha}  links the (constant) angular acceleration to the physical characteristics of the yoyo and the gravitational acceleration when the yoyo is either climbing or falling on the \textit{left} of the string as shown in Fig.~\ref{fig-yoyotradicional}. The $\pm$ reflects the fact the acceleration
depends on the orientation of the yoyo relative to the string.

\begin{figure}[h]
\begin{center}
\includegraphics[width=0.6\textwidth]{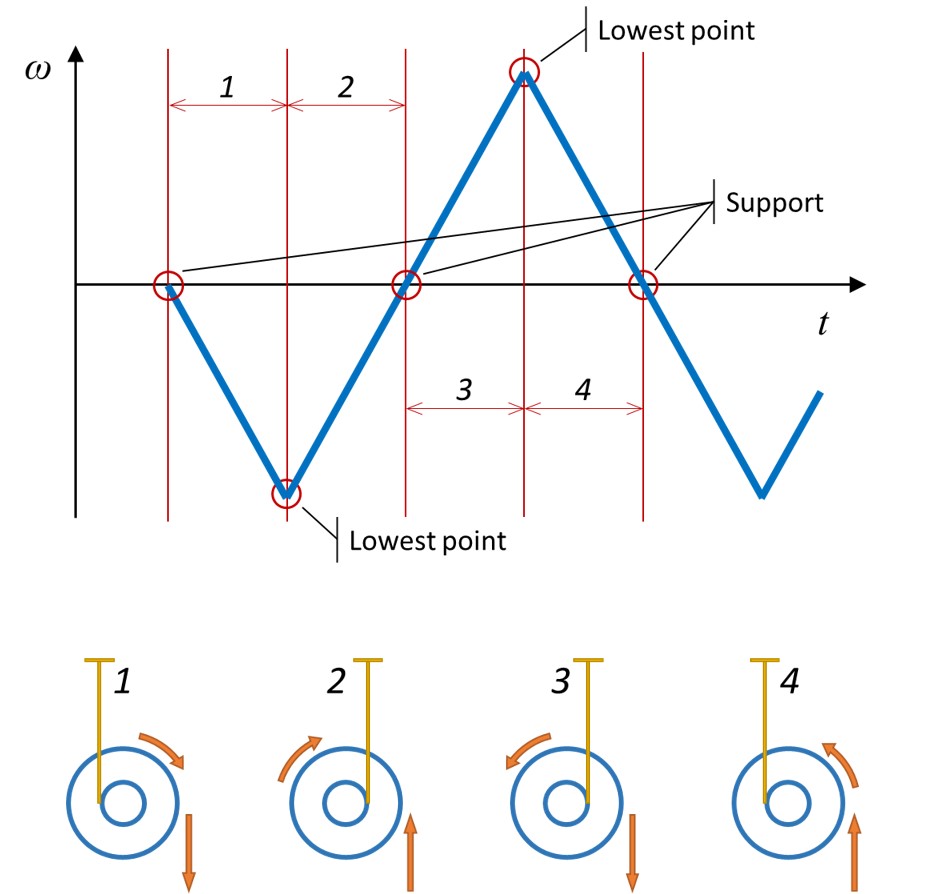}
	\caption{Scheme of the angular velocity  as a function of time  (top panel) and its orientation relative to the string (bottom panel) in each of the time periods ($1,2,3,4$) indicated. The angular velocity displays a saw tooth pattern while the angular acceleration
	(not shown here) takes opposite (constant) values depending whether the yoyo is hanging on the left or right of the string. The changes in the slope occur when the yoyo reaches the lowest position. The arrows indicate the sense of the rotation and the vertical motion. 
	}
\label{fig-evol}
\end{center}
\end{figure}

The expected temporal evolution of the angular velocity deduced from the
previous considerations is depicted in Fig.~\ref{fig-evol}. In the beginning, the yoyo falls along the right side of the string (time period 1). At the lowest point, when the string is fully stretched, the yoyo  behaves like a physical  pendulum during a very brief period of time. In this lapse, the yoyo rotates 180 degrees around the end of the string and the vertical velocity of the center of mass and the angular acceleration change their sign.  As a consequence of the  conservation of angular momentum, the angular velocity when leaving is conserved before and after the yoyo is in the lowest point.   
The origin of the impulsive force applied by the string on the yoyo may be related to the elasticity of the string or  to  sudden displacements of the support (or player's hand).
After that, the yoyo starts climbing,  but now along the \textit{opposite} side of the string, decreasing its angular velocity  (time period 2). When it arrives at the support the angular velocity is null, and the yoyo falls again with the same angular acceleration,
along the left side but now  rotating in the opposite sense (time period 3). Finally, during the last period, the yoyo goes upwards again but on the right side (time period 4).

\begin{figure}
\begin{center}
\includegraphics[width=0.45\textwidth]{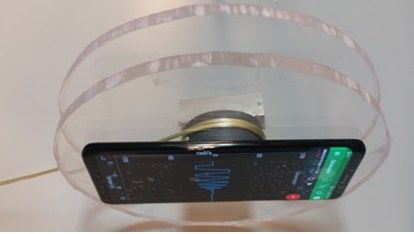}
\caption{Experimental setup: home-made yoyo, counterweight and smartphone.}
\label{fig-yoyosmartphone}
\end{center}
\end{figure}

\begin{figure}
\includegraphics[width=0.73\textwidth]{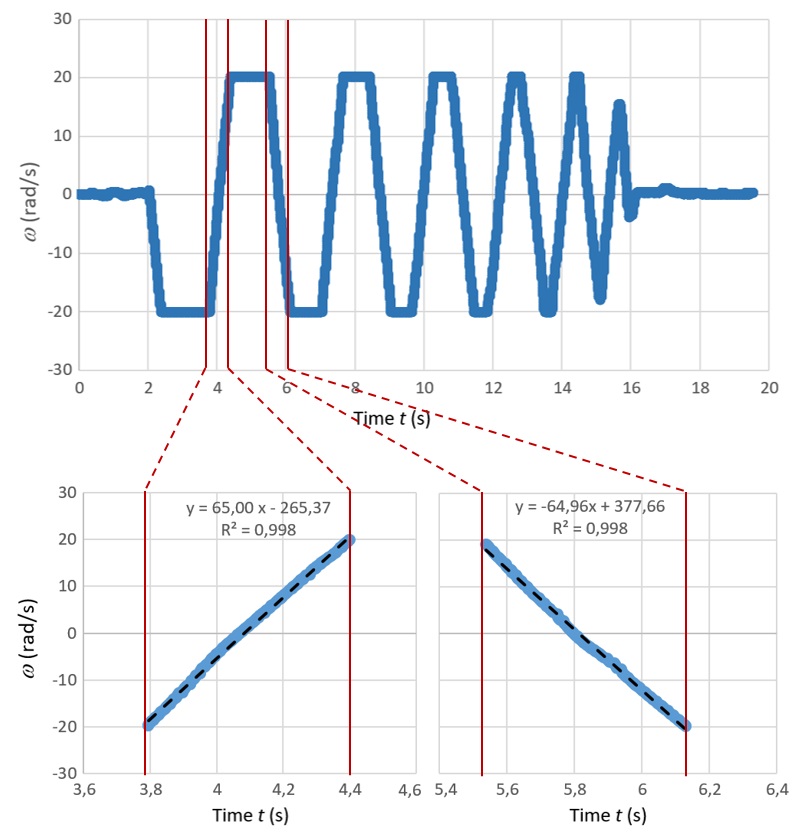}
\caption{Experimental results for the whole temporal evolution of the angular velocity (upper panel)
and a zoom with two windows (bottom panel). In each window the yoyo is climbing and falling along one or the other side of the string.}	
\label{fig-vel-ang-larga}
\end{figure}

\textbf{The experiment.} Our setup comprised a yoyo and a smartphone (Samsung S8+)
with a built-in gyroscope. Since smartphones are usually more voluminous than typical yoyos we built a home-made yoyo which  consists of two methacrylate plates (diameter $19.5$ cm, and thickness $0.8$ cm) joined by an empty  PVC pipe  (external diameter $5.0$ cm and length $2.2$ cm) where a string is wound. In one of the exterior faces of the yoyo a smartphone is fixed with its center of mass coinciding with the yoyo axes, while in the other face a counterweight was placed (Fig.~\ref{fig-yoyosmartphone}).
The Physics Toolbox Suite \textit{app} \cite{physicstoolbox} was used to visualize and register experimental data and export to a CSV (comma separated values) file. A screenshot of the app  displaying the temporal evolution of the angular velocity during the whole movement can be appreciated 
(Fig.~\ref{fig-vel-ang-larga}).

\textbf{Results.} The yoyo is held with the string wound and then released while the smartphone registers the angular velocity along several bounces.  In  Fig.~\ref{fig-vel-ang-larga} we show  the temporal evolution of the  angular velocity during a whole move. At the initial time, the angular velocity is null. When the yoyo is released,  it starts going downwards as the string is unwound and the absolute value of the angular velocity increases. Remarkably, it can be seen that the graph displays several \textit{plateaux} due to the saturation of the sensor  when the angular velocity reaches $20\, rad/s$
(this specific value depends on the smartphone employed).  At successive downwards and upwards travels, the energy is slowly dissipated and the \textit{plateau} disappears.

The overall outlook of the temporal behavior displayed in  Fig.~\ref{fig-vel-ang-larga}
can be understood in the light of the previous discussion. The saw tooth comprises sections
with approximately constant slopes whose (opposite) values
correspond to Eq.\ref{eq:alpha}. In addition, the changes of slope occur when the yoyo reaches the lowest point and the zeros correspond to the passing near the supporting point (or the hand).
In the lower panel, two temporal windows are zoomed in, the first as the yoyo travels up, the second as the yoyo travels down, but, as indicating in the previous paragraph, rotating in opposite directions.

Since the rotation is nearly uniformly accelerated,  to obtain the angular acceleration, we perform a linear fit, $\omega = \omega_0 + \alpha t$, resulting $65.00 \, rad / s^2$ in the first interval (on the
left side of the string) and $-64.96 \, rad / s^2$ in the second (in the right side) showing, 
absolute value, a minimum discrepancy of 0.06\% between both instances.
In addition, it can be noticed, in both temporal windows, a small concavity of the angular velocities. The causes of this phenomenon, to be studied elsewhere,
could be related to several effects such as rolling resistance, 
the non-uniform friction of the thread with the walls of the yoyo or also
the alternation of periods of slipping and rolling related to the static and kinetic friction effects.

\begin{figure}
\includegraphics[width=0.88\textwidth]{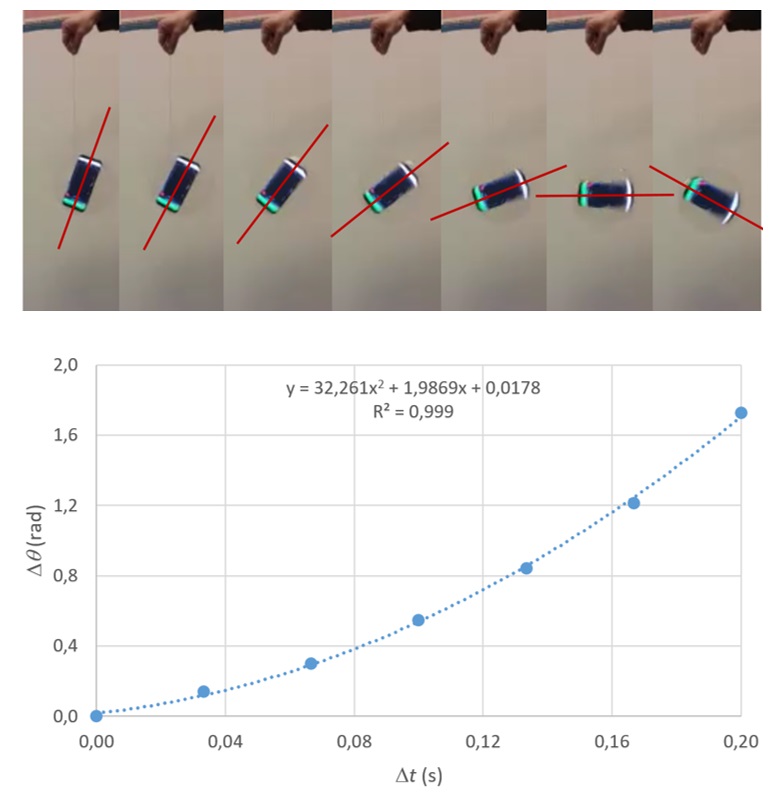}
\caption{The successive screenshots of the smartphone (upper panel) were used to obtain the angular variable $\Delta \theta(t)$  (bottom panel) and the angular acceleration.}
\label{fig-video-analisis}
\end{figure}

Video analysis provides a useful tool to compare with the experimental results obtained from the rotation sensor. The analysis was made from frames extracted from the video at 30 fps. On top of Fig.~\ref{fig-video-analisis} we show 7 consecutive frames in which the yoyo is descending. To determine the angles, we mark the axis of the mobile approximately with a segment and  from its projections the angle with the vertical is obtained. In the lower panel of  Fig.~\ref{fig-video-analisis} we show the angular variable as a function of time. Taking into account the uniformly accelerated circular movement, we perform a quadratic fit
\begin{equation}
\Delta \theta  = \omega_0 \Delta t + \frac 12 \alpha (\Delta t)^2
\end{equation}
which results in an angular acceleration  $ \alpha = 64.52 rad/s^2.$ This value 
appears very close to both previously obtained using the smartphone sensor (mean deviation 
$ 0.7\%$).

\textbf{Conclusion.} In this work we propose a simple experiment using a traditional toy, the yoyo, and a modern device, the smartphone which involves several basics concepts in mechanics.
Thanks to the gyroscope sensor the dynamics of the toy can be accurately analyzed and compared with results obtained from video analysis. The gyroscope sensor provides the angular velocity and, by means of a linear fit, the angular accelerations can be also obtained. We analyzed a whole movement of the yoyo and then focus our attention on two temporal windows. The accelerations in each window are very similar and also coherent with the results obtained analyzing frame by frame the video. It must be emphasized that, depending on the altitude of the thrown, the smartphone sensor could not be able to register all the range of angular velocity values.

One important feature of smartphone sensors to bear in mind
is the coordinate frame relative to they provide measures.
While the accelerometer measures acceleration in a moving frame (relative to the smartphone) the gyroscope measures the angular 
velocity with respect to an inertial reference frame \cite{MONTEIRO2015}. 
In the present experiment, as the smartphone --fixed to the yoyo-- is rotating, the gyroscope is more appropriate than the accelerometer to get a useful measure.

Several aspects of this experiment are worth discussing in a classroom activity. 
An interesting starting point is to present the problem to the students, let them
discuss and predict the evolution of the angular variables and, then, to perform the experiment
and compare the prediction with the results. 
The design and  assembly of the yoyo is the first stage which involves the mass-balance, the selection of the different pieces and the calculation of their contribution to the moment of inertia of the system. The experiment itself requires to set up the sampling
frequency of the sensors and to take into account their maximum range. And finally, the discussion of the results involves all the aspects.
Other possible classrooms proposals go far beyond the experiment reported here and could include the analysis of the mechanical energy, the conservation of angular momentum and  2-dimensional tricks as the described in Ref.\cite{yoyohistory}. We estimate that, 
as this proposal matches classical physics principles and modern technology, it  could be a valuable pedagogical tool to promote students engagement and critical thinking.

We thank the Institute of Educational Sciences of the Universitat Politècnica de València (Spain) for the support of the Teaching Innovation Groups MoMa and e-MACAFI. We also thank the support from the program CSIC \textit{Grupos I+D} and Pedeciba (Uruguay).

\end{document}